# Automatic Short –Answer Grading System (ASAGS)

[1]P.Selvi, [2]Dr.A.K.Bnerjee
[1]Research Scholar, Department of Computer Science and Engineering,
National Institute of Technology, Tiruchirappalli, Tamilnadu, India
p_selvirajendran@yahoo.co.uk
[2]Professor, Department of Computer Science and Engineering,
National Institute of Technology, Tiruchirappalli, Tamilnadu, India

**Abstract –** Automatic assessment needs short answer based evaluation and automated assessment. Various techniques used are Ontology, Semantic similarity matching and Statistical methods. An automatic short answer assessment system is attempted in this paper. Through experiments performed on a data set, we show that the semantic ASAGS outperforms methods based on simple lexical matching; resulting is up to 59 percent with respect to the traditional vector-based similarity metric.

## 1. INTRODUCTION

Automatic assessment is preferred to Manual Assessment to avoid monotonic, bias errors and conserves teacher's time for main activity. Hence automatic assessment is vital for educational system. The area of Computer-based Assessment Systems (CbAS) has grown exponentially due to larger intake by university system, e-learning system as ubiquitous education platform. Computer Assisted Assessment (CAA) is an important area of research due to developments in Natural Language processing (NLP), Information Extraction (IE) and e-learning. [1, 2, 3].

Keyword analysis, full natural-language processing and Information Extraction techniques [2, 4] are the three major techniques for free text assessment system. Keyword analysis has usually been considered a poor method, given that it is difficult to tackle problems such as synonymy or polysemy in the student answers, on the other hand, a full text parsing and semantic analysis is hard to accomplish, and very difficult to port across languages. Hence, Information Extraction offers an affordable and more robust approach, making use of NLP tools for searching the texts for the specific contents and without doing an in-depth analysis [4].

Methods like combining keyword based methods [5], pattern matching techniques [6], breaking the answers into concepts and their semantic dependencies [7], Machine Learning techniques [8], Latent Semantic Analysis (LSA) [9], and LSA with syntactic and semantic information [10, 11] are the other techniques used for the assessment of student's free text answers.

This research envisages the automatic assessment by enhanced BLEU method. The enhanced version of BLEU algorithm is explained in section 2. We describe the system architecture of ASAGS and metrics for evaluating the quality of an automatic scoring algorithm in section 3. The Section 4 illustrates about the experimentation performed on the proposed system.

## 2. THE ENHANCED BLEU METHOD

This method assesses a text by computing a score based on explicit word-to-word match between the student's answer and teacher's answer (i.e. reference). If more than one reference is available, the matching similarity is scored against each reference independently and the best scoring pair is used to find the final score. The unigrams are matched based on the following modules.

**Exact module:** This module will match unigrams only if their surface forms match.

**Stemming module:** This matches two unigrams to each other if they are identical after being passed through the Porter stemmer.

**Heuristics Rule based module:** This module maps two unigrams to each other if they share the same base form based on some heuristics rules.

**Rule 1 - WordNet synonym match:** if two unigrams are matched it shows that they both will have same parts of speech and belongs to the same synset in WordNet.

**Rule 2 - Numeric value match:** The numeric value features to each part of text inferred to correspond to a numeric value. (Eg. "7th"is aligned to "seventh")

**Rule 3 - Acronym match:** It aligns pairs of node with the properties of capitalized letters and the letters correspond to the first characters of some multiword. (Eg. "NLP" is aligned with "Natural Language Processing")

**Rule 4 - Derivational form match:** This Rule is to align words which have the same root form (or have a synonym with the same root form) and which have similar semantic meaning, but which may belong to different syntactic categories.

**Rule 5 - Country adjectival form / demonym match:** It matches from an explicit list of place names, adjectival forms, and demonyms.( Eg. "Chennai" and "Madras")

The steps of Enhanced BLEU [7, 8, 9] algorithm are given below.
1. The matching of N-grams is counted for each reference (1..n). The frequency of each N-gram is clipped with the maximum frequency with which it appears in any reference.
2. Combine the scores of each reference as the weighted linear averages of marks.
3. The short and irrelevant answers are penalized by a penalty factor.

Consequently, sensitivity of procedure hinges on choice of answers.

## 3. SYSTEM ARCHITECTURE OF ASAGS

Figure 1 illustrates the system architecture of ASAGS**.** Various modules used in the system are explained as follows:

**Preprocessing Module:** This module transforms the student's answers into Wraetlic XML format to be processed by the Wraetlic toolkit. The texts are broken into tokens (e.g. words, numbers and punctuation symbols) and the sentence boundaries are identified. The other processes like stemming and stop-words removal are also the part of this module.

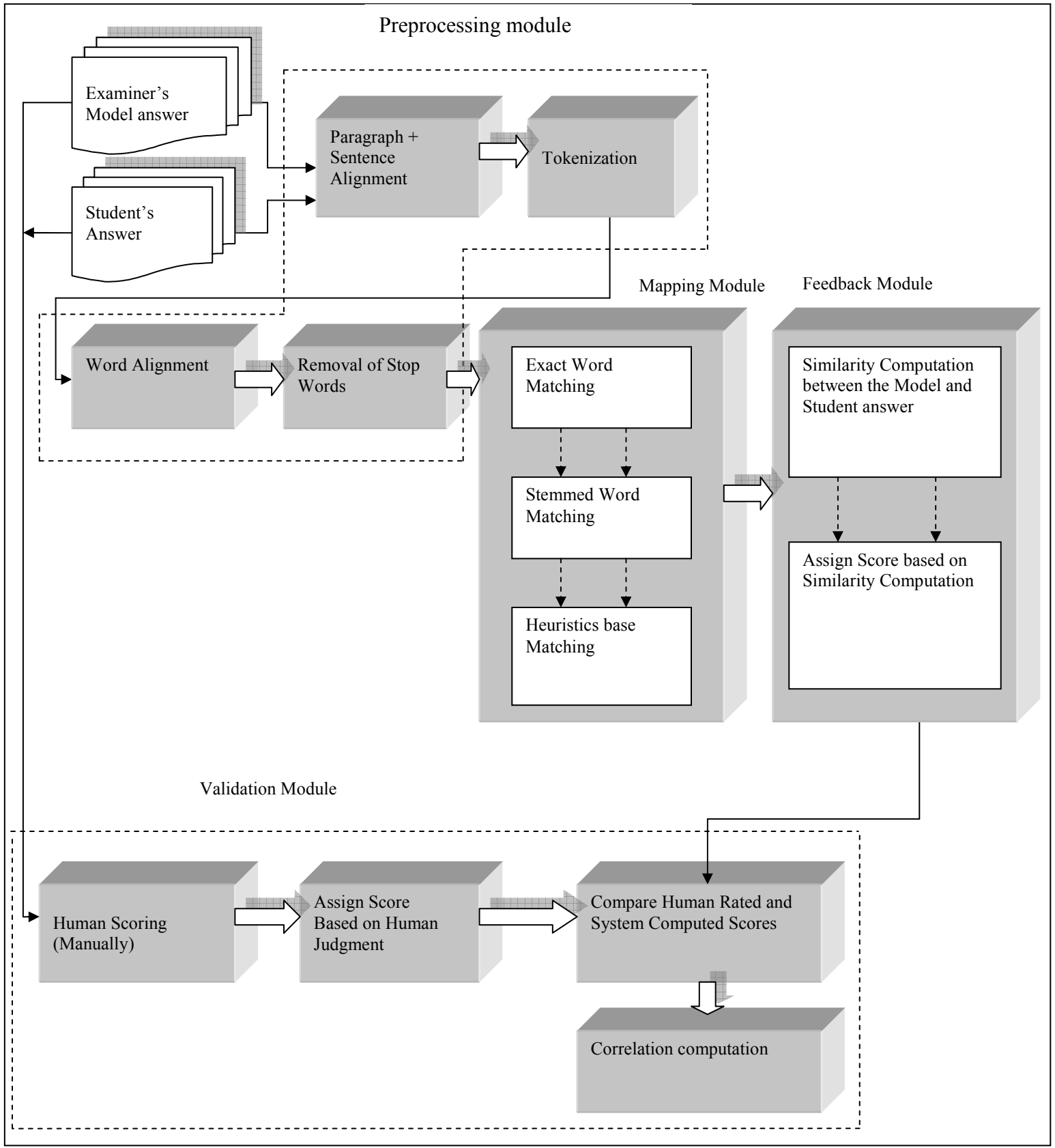

**Figure 1.** Architecture of Automatic Short Answer Grading System (ASAGS)

**Mapping Module:** This module maps the unigrams from the student's answer to the model answers based on the techniques such as exact, stemmed and heuristics rule. As stated before, this module is based on the mapping techniques that can be used independently or together.

**Feedback module:** The system provides effective feedback to students and teachers, the same can be used for training. The numerical score is the result of the comparison module. Its main goal is to provide the student with an orientation about how well she/he has answered according to the numerical scale provided by the teacher.

**Validation Module:** In this module, the data sets that we used for the experimentation are already evaluated by human judges who are experts in the concerned subjects. In comparison steps, the human score and system score are compared and the correlation between human and system score is computed.

## 4. EXPERIMENTATION AND EVALUATION

A bench mark data set released by Rada Mihalcea, Michael Mohler [13] and another created from actual evaluations in our college. The number of students participated range from 14 to 295 based on the question. In particular, three different experiments in the computer science and engineering discipline in our college have been carried out during the 2006-2007 and 2007-2008 academic years:

The ten sets sum up to a total of 1929 student's answers and many different alternative keys were provided and evaluated by different teachers that consisted of descriptions, definitions, Yes or No, advantages and disadvantages as themes. Pearson R (student text vs Reference text) is determined concerning the type of questions, ASAGS works better with convergent questions such as descriptions or definitions in which the sentence order is not so important and they can be compared against reference answers. In fact, three main types of questions have been assessed with ASAGS: definitions, advantages or disadvantages and Yes or No with justification.

The metric used to evaluate the goodness of the free-text scoring of answers to this corpus has been the Pearson correlation filling one vector of scores with the human's scores and the other with the automatic scores.

In this way, the algorithm has been evaluated, for each of the data sets, by comparing the N-grams from the student answers against the references, and obtaining the final score for each candidate. We have varied the following parameters proposed system by [13]:

- N (length of maximum n-gram)
- The measure of recall is taken into account to penalize short answers.

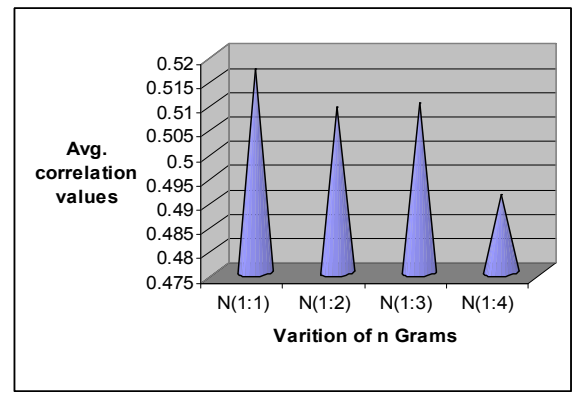

**Figure 2.** Unigram – 4-gram Comparison with Average Correlation Values

We have varied the maximum size of the N-grams taken into consideration from 1 to 4. As Figure 2 considers 1 to 4-gram and shows improvement up to three gram and negative result with respect to 4-gram as students answer having combination of four words match is rare.

### 4.1 Comparison with Other Metrics

The ASAGS system correlation is compared with other metrics and shown in Figure 2

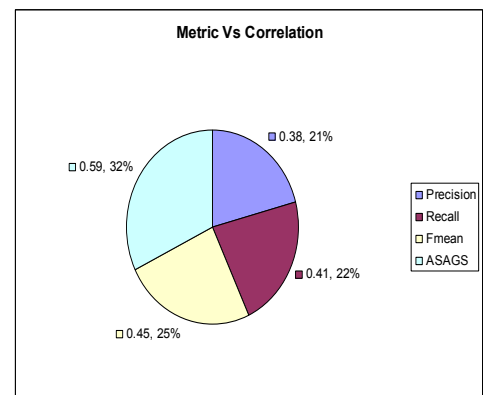

**Figure 3.** Correlation Comparison

We observe in Figure 3 that recall by itself correlates with human assessment much better than precision, and that combining the two using the $F_{mean}$ formula described above results in further improvement. By penalizing the $F_{mean}$ score using the irrelevant response we get some further marginal improvement in correlation.

### 4.2 Different Mapping Comparisons

Effects of various ordering of mapping modules are tested. Table 1 shows the correlation results of different ordering of the mapping modules.

We observe from the table, adding either stemming modules to simply using the exact matching improves correlations. Adding the heuristics based module produces some further improvement in correlation

**Table 1.** Comparing correlations produced by different module stages

| Exp. | Mapping Modules | Correlation |
|---|---|---|
| 1 | Exact only | 0.46 |
| 2 | Exact, Porter Stemmer | 0.48 |
| 3 | Exact, Heuristics | 0.49 |
| 4 | Exact, Porter Stemmer, Heuristics | 0.59 |

### 4.3 Comparison with Existing Evaluation Algorithms

Baseline scoring algorithms used in this work include:

**Keywords**, consists in looking for coincident keywords or n-grams in any of the reference texts but it cannot deal with synonyms or with polysemous terms in the student answers.

**VSM**, using a vectorial representation of similar answers, we have done a five-fold cross-evaluation, in which 20% of the candidate texts are taken as training set for calculating tf.idf weights for each term. Scores for the rest of the answers were through similarities.

**ERB,** The main principle behind ERB algorithm [13] is the measurement of the overlap in unigrams (single words) and higher order n-grams of words, between a student text and a set of one or more reference text.

Table 2 shows that comparison of ASAGS scores with other methods. The first column indicates the scores obtained by ASAGS. The other existing methods ERB, keywords and VSM are represented in consecutive columns. The ASAGS method gives the better score for data set 5. The VSM method obtained the least score. In addition, the VSM method could not evaluate some data sets such as data set 4, 9 and 10. The ASAGS outperforms the other method for 80% of the dataset. It gives the good average correlation compared with other methods.

**Table 2.** Comparison of ASAGS with three other methods

| Data Sets | ASAGS | ERB | Key words | VSM |
|---|---|---|---|---|
| 1 | **0.73** | 0.58 | 0.07 | 0.31 |
| 2 | **0.41** | 0.36 | 0.23 | 0.09 |
| 3 | **0.41** | 0.36 | 0.19 | 0.24 |
| 4 | 0.53 | **0.82** | 0.57 | ----- |
| 5 | 0.46 | 0.41 | **0.57** | 0.52 |
| 6 | **0.09** | 0.02 | -0.05 | 0.05 |
| 7 | **0.41** | 0.21 | 0.32 | 0.17 |
| 8 | **0.51** | 0.41 | 0.22 | 0.17 |
| 9 | **0.81** | 0.73 | 0.24 | ---- |
| 10 | **0.85** | 0.75 | 0.09 | ---- |

The ERB and Keyword based method gives higher correlation for few cases.

### 5. CONCLUSION

Thus we have implemented ASAGS based on heuristics and compared with other methods in literature. Our approach outperforms the other methods for 80% of the dataset and the ERB and Keyword based method for few cases. Other methods like automating Reference text production, and by treatment of synonyms or extending parsing techniques can be probed.